\documentclass[reprint,prl,twocolumn,superscriptaddress,showpacs,showkeys,floatfix,preprintnumbers]{revtex4-1}

\pdfoutput=1

\usepackage{braket}
\usepackage{xargs}
\usepackage{mathtools}
\usepackage[english]{babel}
\usepackage[utf8]{inputenc}
\usepackage{amsmath,amssymb,braket,slashed,mathrsfs}
\usepackage{pifont}
\usepackage{MnSymbol}
\usepackage{graphicx}
\usepackage{tikz}
\usetikzlibrary{shapes,arrows,positioning}
\tikzset{decision/.style={diamond, draw, fill=blue!20, text width=4.5em, text badly centered, inner sep=0pt}}
\tikzset{block/.style={rectangle, draw, fill=blue!20, text width=10em, text centered, rounded corners, minimum width=3.5cm}}
\tikzset{block1/.style={rectangle, draw, fill=blue!20, text width=18.5em, text centered, rounded corners, minimum width=3.5cm}}
\tikzset{line/.style={draw, -latex, thick}}
\usepackage{color,hyperref}
\usepackage{multirow,array}

\usepackage{cleveref}
\newcommand{\ba}{\begin{eqnarray}}
\newcommand{\ea}{\end{eqnarray}}
\newcommand{\be}{\begin{equation}}
\newcommand{\ee}{\end{equation}}

\newcommand{\nn}{\nonumber}

\newcommand{\innovation}{Collaborative Innovation Center of Quantum Matter, Beijing 100871, China}
\newcommand{\chep}{Center for High Energy Physics, Peking University, Beijing 100871, China}
\newcommand{\pkuphy}{School of Physics, Peking University, Beijing 100871,
China}
\newcommand{\KeyLab}{State Key Laboratory of Nuclear Physics and Technology,
Peking University, Beijing 100871, China}
\newcommand{\Uconn}{Department of Physics, University of Connecticut, Storrs, CT 06269, USA}
\newcommand{\RBRC}{RIKEN-BNL Research Center, Brookhaven National Laboratory, Building 510, Upton, NY 11973}

\begin{document}
\title{Light-Neutrino Exchange and Long-Distance Contributions to $0\nu2\beta$
Decays: An Exploratory Study on $\pi\pi\to ee$}

\author{Xu~Feng}\email{xu.feng@pku.edu.cn}\affiliation{\pkuphy}\affiliation{\innovation}\affiliation{\chep}\affiliation{\KeyLab}
\author{Lu-Chang Jin}\email{ljin.luchang@gmail.com}\affiliation{\Uconn}\affiliation{\RBRC}
\author{Xin-Yu Tuo}\affiliation{\pkuphy}
\author{Shi-Cheng Xia}\affiliation{\pkuphy}
%

\date{\today}

\begin{abstract}
We present an exploratory lattice QCD calculation of the neutrinoless double beta
decay $\pi\pi\to ee$. Under the mechanism of light-neutrino exchange, the decay
amplitude involves significant long-distance contributions. 
The calculation reported here, with pion masses $m_\pi=420$ and 140 MeV, demonstrates that
the decay amplitude can be computed from first principles using lattice methods.
At unphysical and physical pion masses,
we obtain that amplitudes are $24\%$ and $9\%$ smaller than the predication from leading order
chiral perturbation theory. Our findings provide the lattice QCD inputs and
constraints
for effective field theory.
A follow-on calculation with fully controlled systematic
errors will be possible with adequate computational resources.
\end{abstract}

\maketitle
   
{\em Introduction.} -- It is a fundamental question whether the neutrinos are Dirac or Majorana-type fermions. 
Neutrinoless double beta ($0\nu2\beta$) decay, if detected, would prove that
neutrinos are Majorana fermions. Besides, it provides direct evidence that the
fundamental law of lepton number conservation is violated in nature. According
to the light-neutrino exchange mechanism, the observation of $0\nu2\beta$ decay would also give us information about the absolute neutrino mass, which oscillation experiments cannot predict.

Around the world many experiments are underway to hunt for $0\nu2\beta$
decays~\cite{Gando:2012zm,Agostini:2013mzu,Albert:2014awa,Andringa:2015tza,KamLAND-Zen:2016pfg,Elliott:2016ble,Agostini:2017iyd,Azzolini:2018dyb,Aalseth:2017btx,Albert:2017owj,Alduino:2017ehq,Agostini:2018tnm}.
Recently four experiments reported the decay's half-lives of
$T_{1/2}^{0\nu}>10^{25}$
yr~\cite{Aalseth:2017btx,Albert:2017owj,Alduino:2017ehq,Agostini:2018tnm} and
a fifth experiment reached the level of $1.07\times10^{26}$ yr for $^{126}$Xe~\cite{KamLAND-Zen:2016pfg}. With a new generation of ton-scale experiments,
the level of sensitivity may be pushed 1 or 2 orders of magnitude higher, yielding the possibility to identify a few decay events per year~\cite{GomezCadenas:2011it,Cremonesi:2013vla,Henning:2016fad,DellOro:2016tmg,Engel:2016xgb}. 

The standard picture of $0\nu2\beta$ involves the long-range light neutrino
exchange -- a minimal extension of the standard model. On the other hand, current knowledge of second-order weak-interaction nuclear matrix elements needs to be improved, as various nuclear models lead to discrepancies on the order of 100\%~\cite{Engel:2016xgb}. 
A promising approach~\cite{Cirigliano:2017tvr,Cirigliano:2018yza} to improving the reliability of the $0\nu2\beta$
predication is to constrain the few-body inputs to {\em ab initio} many-body
calculations using lattice
QCD~\cite{Tiburzi:2017iux,Shanahan:2017bgi,Nicholson:2016byl,Nicholson:2018mwc}.

In this work we perform the first lattice QCD calculation of the nonlocal matrix elements for 
the process of $\pi\pi\to ee$, where the light neutrinos are included as active degrees of freedom. 
We find that the decay amplitude receives dominant long-distance contributions
from the $e\bar{\nu}\pi$ intermediate state. Although small, the excited-state
contribution is identified with a clear signal in our calculation. At both
unphysical and physical pion masses, we find that the lattice results are
consistently smaller than the predication from leading order chiral perturbation
theory~\cite{Cirigliano:2017tvr}.

{\em Light-neutrino exchange in $0\nu2\beta$ decay.} --
We begin with the effective Lagrangian
$\mathcal{L}_{\mathrm{eff}}$ for the single $\beta$ decay
\be
\mathcal{L}_{\mathrm{eff}}=2\sqrt{2}G_FV_{ud}(\bar{u}_L\gamma_\mu
d_L)(\bar{e}_L\gamma_\mu\nu_{eL}),
\ee
which 
represents the standard Fermi charged-current weak interaction involving the
left-handed fermionic fields $\bar{u}_L$, $d_L$, $\bar{e}_L$ and $\nu_{eL}$.
Here $G_F$ is the Fermi constant and $V_{ud}$ is the CKM matrix element. 
One can introduce the neutrino mixing
matrix to connect the neutrino flavor
eigenstates to the mass eigenstates.
For the electron flavor, we have
\be
\bar{e}_L\gamma_\mu\nu_{eL}= \sum_{k=1,2,3}\bar{e}_L\gamma_\mu
U_{ek}\nu_{kL}
\ee
with $U_{ek}$ the mixing matrix element. 

The effective Hamiltonian for $2\beta$ decay can be constructed as
\ba
\mathcal{H}_{\mathrm{eff}}&=&\frac{1}{2!}\int
d^4x\,T\left[\mathcal{L}_{\mathrm{eff}}(x)\mathcal{L}_{\mathrm{eff}}(0)\right]\nn\\
&=&4G_F^2V_{ud}^2
\int d^4x\,H_{\mu\nu}(x)L_{\mu\nu}(x),
\ea
where the hadronic factor $H_{\mu\nu}(x)=T[J_{\mu L}(x)\,J_{\nu L}(0)]$ with
 $J_{\mu L}(x)=\bar{u}_L\gamma_\mu d_L(x)$.
Under the mechanism that $0\nu2\beta$ decays are mediated by the exchange of
light Majorana neutrinos, 
the leptonic factor can be written as~\cite{Doi:1985dx}
\be
L_{\mu\nu}(x)=-m_{\beta\beta}\,S_0(x,0)\overline{e}_L(x)\gamma_\mu\gamma_\nu
e^c_{L}(0)
\ee
with $S_0(x,0)=\int\frac{d^4q}{(2\pi)^4}\frac{e^{iqx}}{q^2}$ a massless scalar
propagator and
$m_{\beta\beta}=\sum_k m_kU_{ek}^2$ the effective neutrino mass. 
The charge conjugate of a fermionic field $\psi$ is given as
$\psi^c=C\bar{\psi}^T=\gamma_4\gamma_2\bar{\psi}^T$.

For a general $0\nu2\beta$ decay $I(p_I)\to F(p_F)e(p_1)e(p_2)$, its decay amplitude can be written as
\ba
\mathcal{A}&=&\langle F,e_1,e_2|\mathcal{H}_{\mathrm{eff}}|I\rangle
\nn\\
&=&-4G_F^2V_{ud}^2m_{\beta\beta}
\int d^4x\,\langle F|H_{\mu\nu}(x)|I\rangle
\nn\\
&&\hspace{0.5cm}\times\int\frac{d^4q}{(2\pi)^4}\,\frac{e^{iqx}}{q^2}\langle e_1,e_2|\overline{e}_L(x)\gamma_\mu\gamma_\nu e^c_{L}(0)|0\rangle.
\ea
Here we use $e_{1,2}$ to specify the electron state carrying momentum $p_{1,2}$.
The leptonic matrix element is given by
$\langle e_1,e_2|\overline{e}_L(x)\gamma_\mu\gamma_\nu e^c_{L}(0)|0\rangle=$
\be
\bar{u}_L(p_1,x)\gamma_\mu\gamma_\nu u^c_L(p_2,0)-\bar{u}_L(p_2,x)\gamma_\mu\gamma_\nu u^c_L(p_1,0),
\ee
which is antisymmetric under the exchange of two electrons $e_1\leftrightarrow e_2$ due to the Pauli exclusion principle.
Here the spinors are defined as
\ba
&&\langle e_i|\bar{e}_L(x)=\bar{u}_L(p_i,x)=\bar{u}_L(p_i)e^{-i\vec{p}_i\cdot\vec{x}}e^{E_it},
\nn\\
&&\langle
e_i|e^c_L(x)=u^c_L(p_i,x)=u^c_L(p_i)e^{-i\vec{p}_i\cdot\vec{x}}e^{E_it},
\ea
for $i=1,2$.
Inserting the complete set of hadronic intermediate states, the decay amplitude can be written as
\ba
&&\mathcal{A}=-4G_F^2V_{ud}^2m_{\beta\beta}\sumint_n\,
\nn\\
&&\hspace{0.2cm}\left[\frac{\langle F|J_{\mu L}|n\rangle\langle
n|J_{\nu L}|I\rangle}{2E_{\nu,2}E_n(E_n+E_{\nu,2}+E_2-E_I)}\bar{u}_L(p_1)\gamma_\mu\gamma_\nu u^c_L(p_2)\right.
\nn\\
&&\hspace{0.2cm}
\left.+\frac{\langle F|J_{\mu L}|n\rangle\langle
n|J_{\nu
L}|I\rangle}{2E_{\nu,1}E_{n}(E_n+E_{\nu,1}+E_1-E_I)}\bar{u}_L(p_1)\gamma_\nu\gamma_\mu
u^c_L(p_2)\right].
\nn\\
\ea
Given the spatial momenta $\vec{p}$ for the hadronic intermediate states
specified by $|n\rangle$, the
neutrino's momenta are constrained by the conservation law
$\vec{p}_{\nu,i}=\vec{p}_I-\vec{p}-\vec{p}_i$ and the corresponding energies
are denoted as $E_{\nu,i}=|\vec{p}_{\nu,i}|$. One can write the spinor product
as a combination of $\bar{u}_L(p_1)u^c_L(p_2)$ and $\bar{u}_L(p_1)\frac{[\gamma_\mu,\gamma_\nu]}{2}u^c_L(p_2)$. The coefficient of the second term is proportional to
the difference in electron momenta and generically suppressed by a factor of $|\vec{p}_1-\vec{p}_2|/k_F\ll 1$, where 
$|\vec{p}_1-\vec{p}_2|\sim O(1)$ MeV and
$k_F\sim O(100)$ MeV is the typical Fermi momentum of nucleons in a nucleus~\cite{Doi:1985dx,Cirigliano:2017tvr}. Keeping only the term of $\bar{u}_L(p_1)u^c_L(p_2)$, the decay amplitude is simplified as
\be
\label{eq:amplitude_simplify}
\mathcal{A}=-T_{\mathrm{lept}}\sumint_n\sum_{i=1,2}\frac{\langle F|J_{\mu L}|n\rangle\langle
n|J_{\mu L}|I\rangle}{2E_{\nu,i}E_n(E_n+E_{\nu,i}+E_i-E_I)}
\ee
with $T_{\mathrm{lept}}=4G_F^2V_{ud}^2m_{\beta\beta}\bar{u}_L(p_1)u^c_L(p_2)$.

{\bf Calculation of $\pi\pi\to ee$ decay} --
In this work we calculate the $\pi\pi\to ee$ decay amplitude with two pions at rest and 
two electrons carrying spatial momenta $\vec{p}_1=-\vec{p}_2$, $|\vec{p}_{1,2}|=E_{\pi\pi}/2$.
While the condition of $|\vec{p}_1-\vec{p}_2|/k_F\ll1$ is no more valid, we
target on the determination of the amplitude given in
Eq.~(\ref{eq:amplitude_simplify}), which is
more relevant for chiral effective field theory
inputs to {\em ab initio} many-body calculation~\cite{Cirigliano:2017tvr}.
This setup has
advantages as follows.
\begin{itemize}
\item Because of the non-zero momentum carried by the electron, the energies of any possible intermediate states $e\bar{\nu} n$ always lie above the initial-state energy $E_{\pi\pi}\approx2m_\pi$. Therefore no exponentially growing contamination is associated with the intermediate states when one performs an integral over a Euclidean time.
The effects of finite volume on a generic second-order weak
    amplitude~\cite{Christ:2015pwa} are not relevant here as well.
\item We use the discrete lattice momenta
$(2\pi/L)\vec{m}$ for the intermediate hadronic particles and the momenta $\vec{p}_{\nu,i}=-\vec{p}_i-(2\pi/L)\vec{m}$ for the intermediate neutrino, where $\vec{p}_i$ is the momentum carried by the electron. As nonzero momenta are assigned 
for the neutrino propagator, one can keep the lowest mode of the propagator,
which reduces the power-law finite-volume effects.
\end{itemize}

Note that no short-distance divergence appears as $x$ approaches to $0$ in $\mathcal{L}_{\mathrm{eff}}(x)\mathcal{L}_{\mathrm{eff}}(0)$. This can be seen by the power counting in the integral
\ba
&&\int d^4x\, e^{i\Lambda x}\mathcal{L}_{\mathrm{eff}}(x)\mathcal{L}_{\mathrm{eff}}(0)
\nn\\
&&\hspace{1cm}\sim 8G_F^2V_{ud}^2
\frac{m_{\beta\beta}}{\Lambda^2}(\bar{u}_L\gamma_\mu d_L)(\bar{u}_L\gamma_\mu d_L)\bar{e}_Le^c_L.
\ea
In lattice QCD, a hard cutoff is introduced by the inverse of lattice spacing $1/a$. Thus the unphysical short-distance contribution appears as an $O(a^2)$ discretization effect.

Using the Coulomb gauge fixed wall sources for the $\phi_\pi$ interpolating operator, we
construct the correlation function through $C(t_x,t_y,t_{\pi\pi})=$
\ba
&=&-4G_F^2V_{ud}^2m_{\beta\beta}\times
\nn\\
&&\Big(\sum_{\vec{x},\vec{y}}\langle 0|T[J_{\mu L}(t_x,\vec{x})J_{\mu L}(t_y,\vec{y})\phi_\pi^\dagger\phi_\pi^\dagger(t_{\pi\pi})]|0\rangle
\nn\\
&&\hspace{1cm}\times S_0(x,y)
\langle e_1e_2|\bar{e}_L(x)e^c_L(y)|0\rangle\Big)
\nn\\
&=&-T_{\mathrm{lept}}\sum_{\vec{x},\vec{y}}\langle 0|T[J_{\mu L}(t_x,\vec{x})J_{\mu L}(t_y,\vec{y})\phi_\pi^\dagger\phi_\pi^\dagger(t_{\pi\pi})]|0\rangle
\nn\\
&&\hspace{1cm}\times S_0(x,y)\left(e^{-i\vec{p}_1\cdot(\vec{x}-\vec{y})}+e^{-i\vec{p}_2\cdot(\vec{x}-\vec{y})}\right)e^{\frac{E_{\pi\pi}}{2}(t_x+t_y)}.
\nn\\
\ea
On the lattice, the scalar propagator $S_0(x,y)e^{-i\vec{k}\cdot(\vec{x}-\vec{y})}$ with $\vec{k}=\vec{p}_{1,2}$ can be implemented as
\ba
&&S_0(x,y)e^{-i\vec{k}\cdot(\vec{x}-\vec{y})}=\int\frac{d^4q}{(2\pi)^4}\frac{e^{iq(x-y)}}{q_t^2+(\vec{q}+\vec{k})^2}
\nn\\
&&\hspace{2cm}\Rightarrow\quad\quad 
\frac{1}{VT}\sum_{\vec{q},q_t}\frac{e^{iq(x-y)}}{\widehat{q_t}^2+\sum_i\widehat{q_i+k_i}^2}
\ea
with $\widehat{q_i}=2\sin(q_i/2)$ the lattice discretized momenta. $V$ and $T$ are the spatial volume and time extent of the lattice. We can calculate the zero mode ($\vec{q}=0$) of the propagator
as $\frac{1}{VT}\sum_{q_t}\frac{e^{iq_t(t_x-t_y)}}{\widehat{q_t}^2+\sum_i\widehat{k_i}^2}$. The nonzero modes ($\vec{q}\neq\vec{0}$) of 
the propagator can be constructed as
$\frac{1}{N_r}\sum_{r=1}^{N_r}\phi_r(x)\phi_r^*(y)$ using the stochastic method,
with
\be
\phi_r(x)=
\frac{1}{\sqrt{VT}}\sum_{\vec{q}\neq\vec{0},q_t}\frac{\xi_r(q)e^{iqx}}{\sqrt{\widehat{q_t}^2+\sum_i\widehat{q_i+k_i}^2}}.
\ee
Here the stochastic sources $\xi_r(q)$ satisfy 
\be
\lim_{N_r\to\infty}\frac{1}{N_r}\sum_r\xi_r(q)\xi_r^*(q')=\delta_{q,q'}.
\ee
It is proposed by the NPLQCD Collaboration that the neutrino
propagator can also be computed in an exact way by using double Fourier
transformation~\cite{Detmold:2018zan}.

Following
Refs.~\cite{Christ:2012se,Bai:2014cva,Christ:2015aha,Christ:2016eae,Christ:2016mmq,Bai:2017fkh,Bai:2018hqu} and integrating $t_x$ and $t_y$ over a fixed window $[t_a,t_b]$ with $t_a\gg t_{\pi\pi}$, we obtain
\ba
\label{eq:integrated_amplitude}
\mathcal{M}&=&\sum_{t_x=t_a}^{t_b}\sum_{t_y=t_a}^{t_b}C(t_x,t_y,t_{\pi\pi})/\left(V\frac{N_{\pi\pi}}{2E_{\pi\pi}}e^{E_{\pi\pi}t_{\pi\pi}}\right)
\nn\\
&=&-T_{\mathrm{lept}}\sum_n\frac{1}{V}\sum_{\vec{p}_n}\sum_{i=1,2}\frac{\langle 0|J_{\mu L}|n\rangle\langle
n|J_{\mu L}|\pi\pi\rangle}{2E_{\nu,i}E_n(E_n+E_{\nu,i}+E_i-E_{\pi\pi})}
\nn\\
&&\hspace{0.8cm}\times\left(T_{\mathrm{box}}+\frac{e^{-(E_n+E_{\nu,i}+E_i-E_{\pi\pi})T_{\mathrm{box}}}-1}{E_n+E_{\nu,i}+E_i-E_{\pi\pi}}\right)
\ea
with $T_{\mathrm{box}}=t_b-t_a+1$ the time extent of the integration window. $N_{\pi\pi}$ and $E_{\pi\pi}$ are known
from the correlation function
$\langle\phi_\pi\phi_\pi(t)\phi_\pi^\dagger\phi_\pi^\dagger(0)\rangle\xrightarrow[]{t\gg0}V\frac{N_{\pi\pi}^2}{2E_{\pi\pi}}\left(e^{-E_{\pi\pi}t}+e^{-E_{\pi\pi}(T-t)}\right)+\mbox{const}$
by using the methods proposed in Ref.~\cite{Feng:2009ij}. When $T_{\mathrm{box}}$ is 
sufficiently large, the contamination from the exponential term vanishes as $E_n+E_{\nu,i}+E_i>E_{\pi\pi}$.
The coefficient of the term proportional to $T_{\mathrm{box}}$ provides a result
for the decay amplitude $\mathcal{A}(\pi\pi\to ee)$.

{\em Numerical results.} --
We use two ensembles with $m_\pi=420$ and 140 MeV generated by the RBC and UKQCD
Collaborations~\cite{Blum:2011pu}. The corresponding parameters are listed in
Table~\ref{tab:ensemble_parameter}. We produce wall-source light-quark propagators on all time slices and make use of the time translation
invariance to average the correlator over all $T$ time translations. (To reduce
the computational costs at $m_\pi=140$ MeV, we adopt the technique of all mode
average~\cite{Collins:2007mh,Blum:2012uh} with $T$ sloppy propagators used for
correlator average and $1$
precise propagator for correction.) We compute propagators for both periodic and 
antiperiodic boundary conditions in the temporal direction and use their average 
in the calculation, which effectively doubles the temporal extent of the lattice.

\begin{table}[htb]
\begin{tabular}{ccccc}
\hline\hline
$m_\pi$ [MeV] & $a^{-1}$ [GeV] & $L^3\times T$ & $N_{\mathrm{conf}}$ & $N_r$  \\
\hline
420 &  1.73 & $16^3\times32$ & 200 & 32 \\
140 &  1.01 & $24^3\times64$ & 60 & 64 \\
\hline
\end{tabular}
\caption{Ensembles used in this work. We list the pion mass $m_\pi$, the lattice
    spacing inverse $a^{-1}$, the space-time volume $L^3\times T$, the number,
    $N_{\mathrm{conf}}$, of
    configurations used and the number, $N_r$, of stochastic sources for the
    neutrino propagator.}
\label{tab:ensemble_parameter}
\end{table}

\begin{figure}
\centering
\includegraphics[width=.2\textwidth]{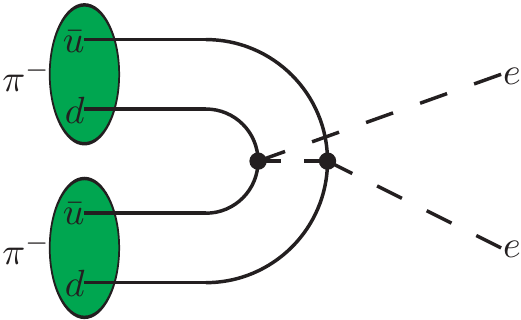}
\hspace{1cm}
\includegraphics[width=.2\textwidth]{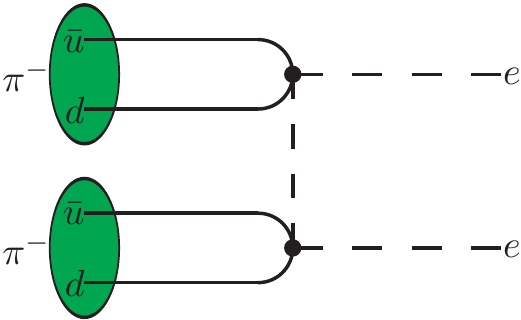}
\caption{Quark and lepton contractions for the process of $\pi\pi\to ee$.}
   \label{fig:contraction}
\end{figure}

The Feynman diagrams corresponding to the process of $\pi\pi\to ee$ are shown in
Fig.~\ref{fig:contraction}. To show the time dependence of the
$C(t_x,t_y,t_{\pi\pi})$ explicitly, we define the unintegrated amplitude
$\mathcal{M}(t)$ as a function of the variable $t=t_x-t_y$:
\be
\label{eq:M_t}
\mathcal{M}(t)=C(t_x,t_y,t_{\pi\pi})/\left(V\frac{N_{\pi\pi}}{2E_{\pi\pi}}e^{E_{\pi\pi}t_{\pi\pi}}\right)
\ee
The time $t_x$ and $t_y$ are separated by at least 6 time units from the $\pi\pi$ sources ($t_{x,y}-t_{\pi\pi}\ge6$) so that the $\phi_\pi^\dagger\phi_\pi^\dagger$ interpolating operators can project onto the ground $\pi\pi$ state. At large $|t|$, the time dependence of $\mathcal{M}(t)$ is saturated by the ground intermediate state - $e\bar{\nu}\pi$
\be
\label{eq:ground_state}
\mathcal{M}(t)\xrightarrow[]{|t|\gg0}-T_{\mathrm{lept}}\frac{1}{V}\frac{2\langle0|J_{\mu
L}|\pi\rangle_{V}\langle\pi|J_{\mu L}|\pi\pi\rangle_{V}}{(2m_\pi)(2E_\nu)}e^{-m_\pi |t|},
\ee
where the matrix elements of $\langle0|J_{\mu L}|\pi\rangle_{V}$ and
$\langle\pi|J_{\mu L}|\pi\pi\rangle_{V}$ are determined from
the correlation functions $\langle J_{\mu L}(t)\phi_\pi^\dagger(0)\rangle$ and
$\langle\phi_\pi(t_\pi)J_{\mu
L}(t_J)\phi_\pi^\dagger\phi_\pi^\dagger(t_{\pi\pi})\rangle$, respectively. The
subscript $\langle\cdots\rangle_{V}$ indicates that
the initial and final states are defined in the finite volume. The single-pion
states $|\pi\rangle_{V}$ satisfy the normalization condition
$\langle\pi(\vec{p})|\pi(\vec{p}')\rangle_{\mathrm{FV}}=(2E_\pi)V\delta_{\vec{p},\vec{p}'}$, while 
the two-pion states $|\pi\pi\rangle_{V}$ can be connected to the states in the finite volume $|\pi\pi\rangle_\infty$ through
the Lellouch-L\"uscher relation~\cite{Lellouch:2000pv,Lin:2001ek}
\be
|\pi\pi\rangle_\infty=\left(2\pi\frac{E_{\pi\pi}}{k^3}\right)^{\frac{1}{2}}\left(q\frac{d\phi}{dq}+k\frac{d\delta}{dk}\right)^{\frac{1}{2}}|\pi\pi\rangle_{V}
\ee
with the momenta $k=\sqrt{\frac{E_{\pi\pi}^2}{4}-m_\pi^2}$ and $q=kL/(2\pi)$.

    \begin{figure}
    \centering
    \includegraphics[width=.48\textwidth]{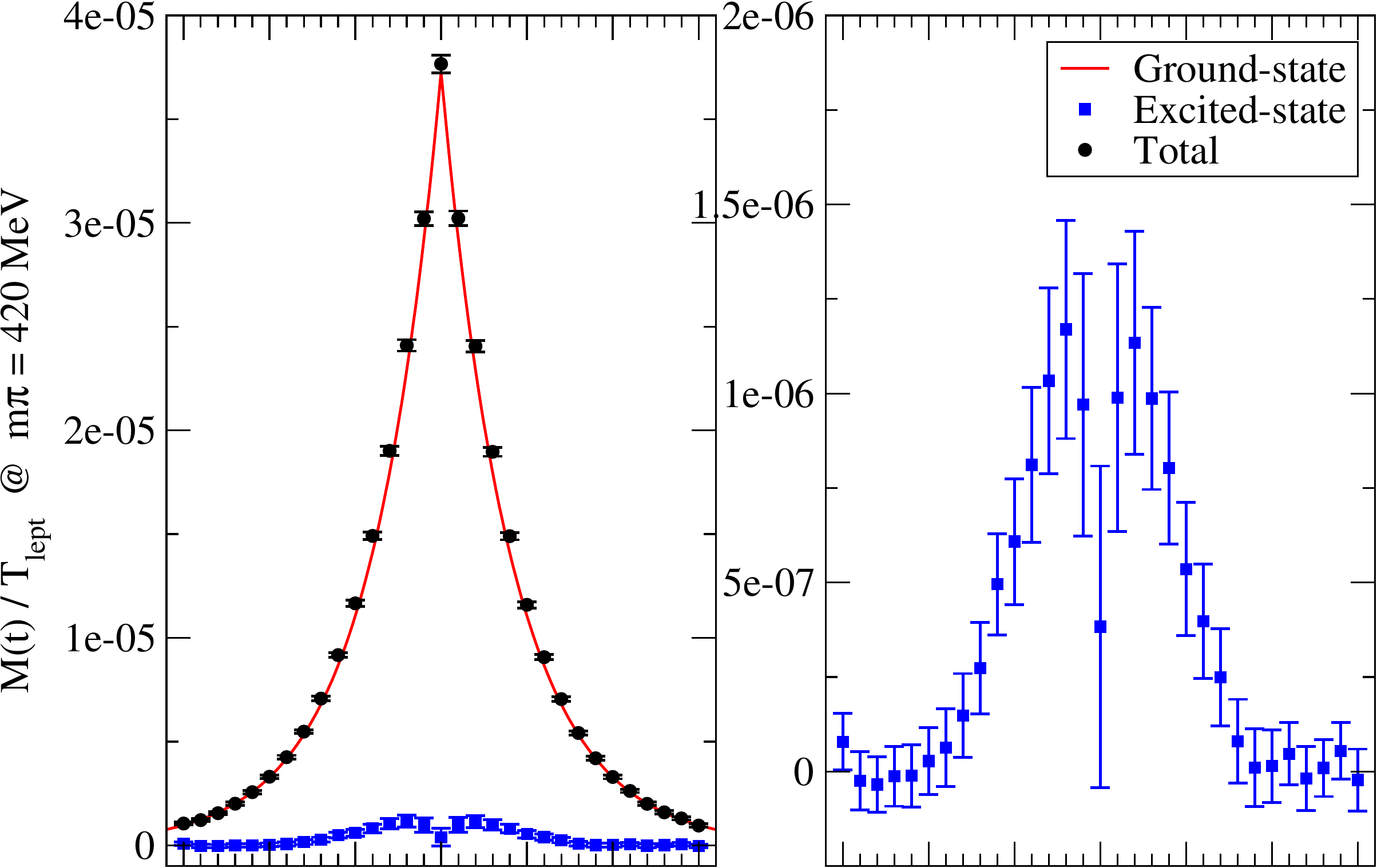}
    \includegraphics[width=.48\textwidth]{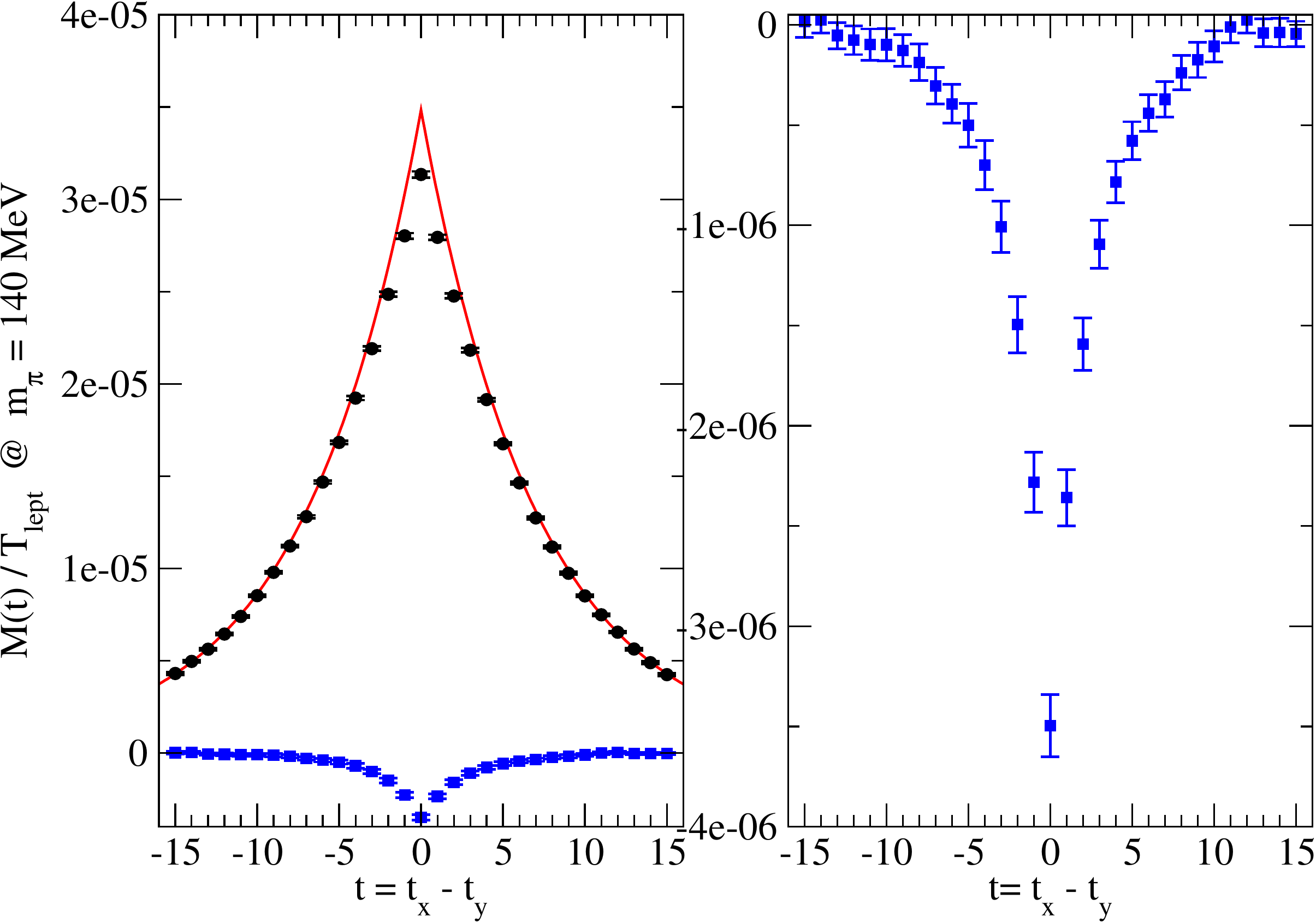}
    \caption{Unintegrated amplitude $\mathcal{M}(t)$ defined in
    Eq.~(\ref{eq:M_t}) as a function of $t=t_x-t_y$. The black circles show the total contribution of
$\mathcal{M}(t)$. The red curve is not a fit to $\mathcal{M}(t)$, but a ground-state contribution predicted by Eq.~(\ref{eq:ground_state}). The blue squares show the remaining excited-state contribution. }
    \label{fig:unintegrated}
    \end{figure}

The time dependence of $\mathcal{M}(t)$ is shown in Fig.~\ref{fig:unintegrated}.
At large $|t|$ the data of $\mathcal{M}(t)$ are consistent with the contribution
from the ground intermediate state. 
By subtracting the ground-state
contribution, the remaining excited-state contribution is shown by blue square
points in the left panel of Fig.~\ref{fig:unintegrated} and enlarged in the
right panel. Although relatively small, the contribution from the excited intermediate states
can be identified with a clear signal.

    \begin{figure}
    \centering
    \includegraphics[width=.48\textwidth]{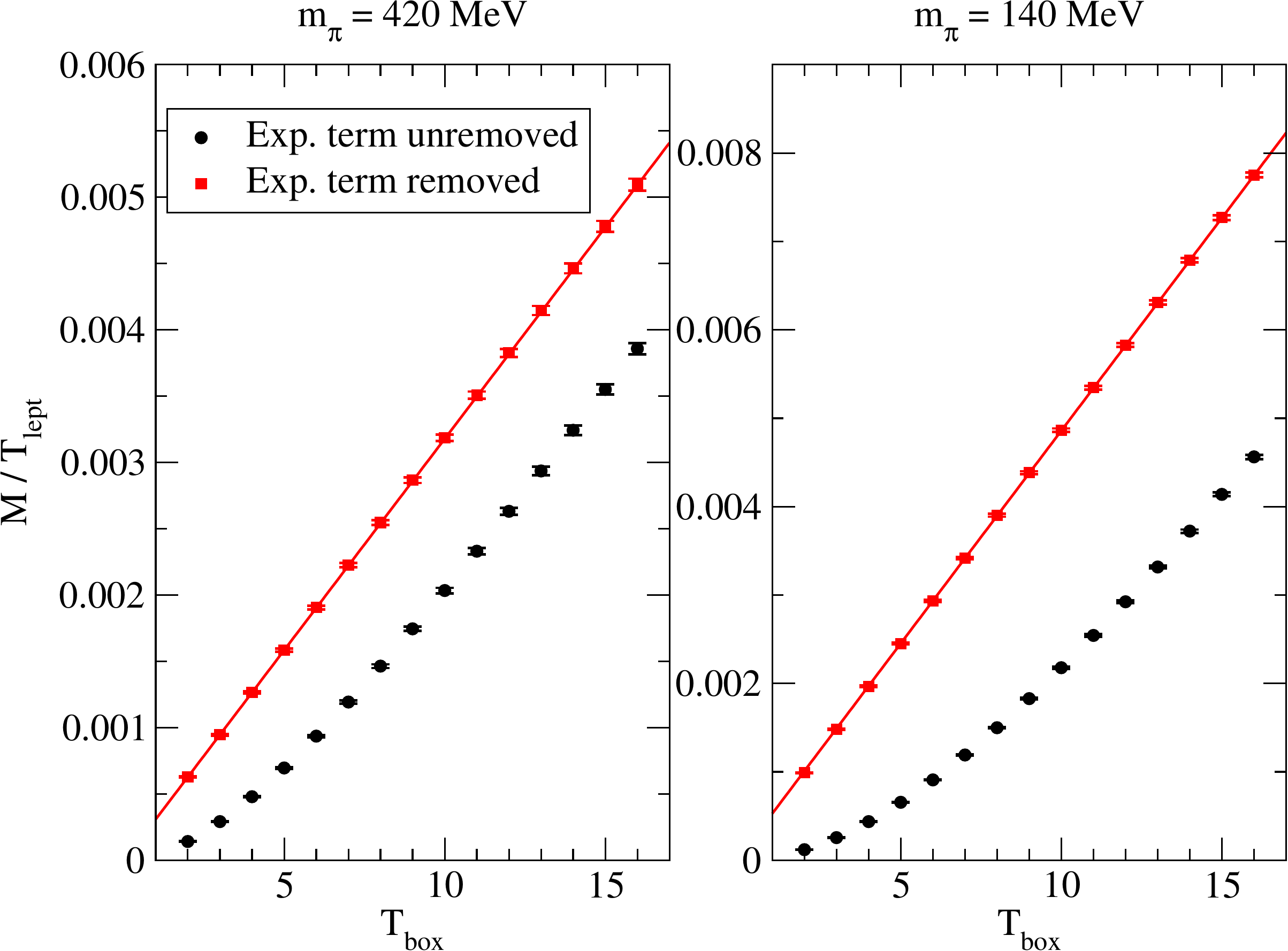}
    \caption{Integrated matrix element $\mathcal{M}$ as a function of $T_{\mathrm{box}}$. The black circles show
the integrated matrix element $\mathcal{M}$ defined in Eq.~(\ref{eq:integrated_amplitude}). The red squares show the results of $\mathcal{M}$
with the exponential term for the ground intermediate state, $\frac{e^{-m_\pi T_{\mathrm{box}}}-1}{m_\pi}$, 
subtracted.}
    \label{fig:integrated}
    \end{figure}

The integrated matrix element defined in Eq.~(\ref{eq:integrated_amplitude}) is
shown in Fig.~\ref{fig:integrated}. We realize that the size of integration
window $T_{\mathrm{box}}\approx 16$ is not sufficiently large to discard the
exponential term associated with the ground intermediate state. (This can be
confirmed in Fig.~\ref{fig:unintegrated} that at $|t|\approx 15$ the values
of $\mathcal{M}(t)$ are statistically larger than $0$.) 
After removing this exponential term, we can fit the lattice data to a linear
function of $T_\mathrm{box}$ and determine the values of $\mathcal{A}_{\mathrm{lat}}(\pi\pi\to ee)$. 
To convert $A_{\mathrm{lat}}(\pi\pi\to ee)$ to the physical amplitude
$\mathcal{A}(\pi\pi\to ee)$, a renormalization factor square $Z_{V/A}^2$ shall be
multiplied, which relates the local lattice vector or axial-vector current
(which we use) to the conserved or partially conserved ones. Besides, the
Lellouch-L\"uscher factor shall be multiplied to relate a finite-volume
amplitude to the infinite-volume one. In our calculation, the two pions are in
the ground state, i.e. at threshold. The large-$L$ expansion of the Lellouch-L\"uscher factor is given by
\ba
&&\frac{2\pi}{k^3}\left(q\frac{d\phi}{dq}+k\frac{d\delta}{dk}\right)
=V\left[1+d_1\frac{a_{\pi\pi}}{L}+d_2\left(\frac{a_{\pi\pi}}{L}\right)^2\right.
\nn\\
&&\hspace{1.8cm}\left.+d_3\left(\frac{a_{\pi\pi}}{L}\right)^3-2\pi\frac{a_{\pi\pi}^2 r_{\pi\pi}}{L^3}+O(L^{-4})\right]
\ea 
with $a_{\pi\pi}$ the scattering length and $r_{\pi\pi}$ the effective range from the $k$ expansion of
$\pi\pi$ scattering phase shift
$k\cot\delta(k)=a_{\pi\pi}^{-1}+r_{\pi\pi}\frac{k^2}{2}+O(k^4)$.
The coefficients $d_i$ are given by
\ba
&&d_1=-2\frac{Z_{00}(1;0)}{\pi}=5.674\,595,
\nn\\
&&d_2=\frac{Z_{00}(1;0)^2+3Z_{00}(2;0)}{\pi^2}=13.075\,478,
\nn\\
&&d_3=\frac{4\pi^4-4Z_{00}(3;0)}{\pi^3}=11.482\,471.
\ea
The values of the zeta function $Z_{00}(s,0)$ have been provided by
Ref.~\cite{Luscher:1986pf}.
We evaluate $a_{\pi\pi}$ using L\"uscher’s
finite-size formula~\cite{Luscher:1986pf} and use it as an input to determine
the finite-volume correction up to $O(L^{-2})$.

Another type of power-law finite-volume effect arises from the long-range
property of the neutrino propagator. 
The finite-volume effects relevant for the
$e\bar{\nu}\pi$-intermediate state can be evaluated
as
\be
\Delta_{\mathrm{FV}}=\left(\frac{1}{V}\sum_{\vec{p}}-\int\frac{d^3\vec{p}}{(2\pi)^3}\right)\frac{\langle0|J_{\mu
     L}|\pi(\vec{p})\rangle\langle\pi(\vec{p})|J_{\mu
    L}|\pi\pi\rangle}{E_\nu E_\pi(E_\pi+E_\nu+E_e-E_{\pi\pi})}
\ee
with $\vec{p}=\frac{2\pi}{L}\vec{n}$ the discrete momentum for the pion.
The neutrino's energy is given by $E_\nu=|\vec{p}+\vec{p}_e|$, with $\vec{p}_e$
the momentum carried by the electron.
We define a function
$f(\vec{p})\equiv\frac{\langle0|J_{\mu
       L}|\pi(\vec{p})\rangle\langle\pi(\vec{p})|J_{\mu
   L}|\pi\pi\rangle}{E_\pi(E_\pi+E_\nu+E_e-E_{\pi\pi})}$
   and split it as $f(\vec{p})=f(-\vec{p}_e)+[f(\vec{p})-f(-\vec{p}_e)]$.
   The term inside brackets does not contribute a power-law finite-volume
   effect. We thus simplify $\Delta_{\mathrm{FV}}$ as
\ba
\label{eq:FV}
\Delta_{\mathrm{FV}}&=&f(-\vec{p}_e)\left(\frac{1}{V}\sum_{\vec{p}}-\int\frac{d^3\vec{p}}{(2\pi)^3}\right)\frac{1}{|\vec{p}_e+\vec{p}|}
\nn\\
&=&f(-\vec{p}_e)\left[-\frac{\kappa(\vec{n}_e)}{2\pi L^2}\right].
\ea
The function $\kappa(\vec{n}_e)$ with
$\vec{n}_e=\vec{p}_eL/(2\pi)$ can be computed numerically and
we find $\kappa(\vec{n}_e)=0.686(3)$ for $m_\pi=420$ MeV and $0.517(3)$ for
$m_\pi=140$ MeV. 
Thus Eq.~(\ref{eq:FV}) indicates that the finite-volume correction appears as an
$O(L^{-2})$ effect.
We expect that the size of $f(-\vec{p}_e)$ is significantly smaller
than $f(\vec{0})$, as the total contribution to the decay amplitude from the 
intermediate hadronic states that carry nonzero lattice momenta
only amounts for 3\%-4\% when compared to the
zero-momentum
contribution. 
We therefore neglect this finite-volume effect in this work, and leave it
for future studies.

In Table~\ref{tab:decay_amplitude}, we show the ground-state, excited-state and
total contributions to the decay amplitude as $\mathcal{A}^{(g)}$,
$\mathcal{A}^{(e)}$ and $\mathcal{A}^{(g)}+\mathcal{A}^{(e)}$, respectively. The
results are presented in units of $F_\pi^2\,T_{\mathrm{lept}}$, where the decay
constant $F_\pi$ is determined from the matrix element $\langle
0|\bar{d}\gamma_\mu\gamma_5u|\pi(p)\rangle=\sqrt{2}p_\mu Z_A F_\pi$, with $Z_A$
the renormalization constant.
Systematic effects associated with three choices of $t_a-t_{\pi\pi}=6,7,8$ are relatively smaller than the statistical errors, suggesting that a separation of 6 is a safe choice to neglect the excited $\pi\pi$ states.

\begin{table}[htb]
\begin{tabular}{ccrrr}
\hline\hline
$m_\pi$ [MeV] & $t_a-t_{\pi\pi}$ & \multicolumn{1}{c}{$\mathcal{A}^{(g)}$} &
\multicolumn{1}{c}{$\mathcal{A}^{(e)}$} &
\multicolumn{1}{c}{$\mathcal{A}^{(g)}+\mathcal{A}^{(e)}$}  \\
\hline
    & 6 &   & $0.055(13)$ & $1.517(13)$ \\
420 & 7 &  $1.462(10)$ & $0.060(13)$ & $1.522(13)$ \\
    & 8 &   & $0.052(14)$ & $1.514(14)$ \\
\hline
    & 6 &  & $-0.0664(70)$ & $1.8200(63)$ \\
140 & 7 & $1.8864(50)$ & $-0.0660(73)$ & $1.8204(62)$ \\
    & 8 &  & $-0.0665(70)$ & $1.8199(60)$ \\
\hline
\end{tabular}
\caption{Results for ground-state ($\mathcal{A}^{(g)}$), excited-state
($\mathcal{A}^{(e)}$) and total  ($\mathcal{A}^{(g)}+\mathcal{A}^{(e)}$)
contributions to the $\pi\pi\to ee$ decay amplitude.
All the results are listed in units of $F_\pi^2 \,T_{\mathrm{lept}}$.}
\label{tab:decay_amplitude}
\end{table}

{\em Conclusion.} --
We have carried out a lattice QCD calculation of the decay
amplitude of $\pi\pi\to ee$ and obtained the result with subpercent statistical
errors:
\ba
&&\frac{\mathcal{A}(\pi\pi\to ee)}{F_\pi^2\,
T_{\mathrm{lept}}}\bigg|_{m_\pi=420\,\,
\mathrm{MeV}}=1.517(13),
\nn\\
&&
\frac{\mathcal{A}(\pi\pi\to ee)}{F_\pi^2\,
T_{\mathrm{lept}}}\bigg|_{m_\pi=140\,\,
\mathrm{MeV}}=1.820(6).
\ea
The decay amplitude of $\mathcal{A}(\pi\pi\to ee)$ is mainly contributed by the ground intermediate state via the process of
$\pi\pi\to\pi e\bar{\nu}\to ee$. Although the size of the excited-state contribution is only
3\%-4\%, it is statistically significant (see Fig.~\ref{fig:unintegrated}) as the
uncertainty of the amplitude has been reduced to below 1\%. 

Without the signal-to-noise problem, the case of $\pi\pi\to ee$ serves as an ideal laboratory to develop the necessary methods and tools for
a calculation of $0\nu2\beta$ decay with controlled uncertainties. 
Our exploratory study demonstrates the possibility of a first-principles calculation of the long-distance contribution to $0\nu2\beta$ decay via light-neutrino exchange. 
At $m_\pi=420$ and 140 MeV, we find that the decay amplitude $\mathcal{A}(\pi\pi\to ee)$ are 24\% and 9\% smaller than the
leading-order predication $\mathcal{A}^{\mathrm{LO}}(\pi\pi\to ee)=2\, F_\pi^2\,
T_{\mathrm{lept}}$ in chiral perturbation theory~\cite{Cirigliano:2017tvr}. 
Various systematic effects such as lattice artifacts and finite-volume effects
require an accurate examination in future work but are not expected to
qualitatively alter the conclusions of this work. The 9\% deviation 
found here is still quite consistent with power counting in
effective field theory. On the other hand, Ref.~\cite{Cirigliano:2018hja} has found that a
leading-order,
short-range contribution needs to be introduced in the $nn\to pp ee$ decay, which 
breaks down Weinberg’s power-counting scheme.
It is interesting to examine the impact of this short-range contribution in our future study.
The techniques presented here can be directly applied to the study of other $0\nu2\beta$ decays, such as $n\pi\to pee$ and $nn\to pp ee$.
From these decays, lattice QCD can provide more low-energy QCD inputs for the
effective field theory~\cite{Cirigliano:2017tvr}.

\begin{acknowledgments}

We gratefully acknowledge many helpful discussions with our colleagues from the
RBC-UKQCD Collaboration. X.F. warmly thanks N.~H.~Christ, W.~Dekens, W. Detmold,
E.~Mereghetti, D.~Murphy and U.~van Kolck for useful discussion. X.F., X.-Y.T.
and S.-C.X. were supported in part by NSFC of China under Grant No. 11775002.
The calculation was carried out on TianHe-1 (A) at Chinese National Supercomputer Center in Tianjin.
Part of the computation was performed under the ALCC Program of the U.S. DOE 
on the Blue Gene/Q (BG/Q) Mira computer at the Argonne Leadership 
Class Facility, a DOE Office of Science Facility supported under Contract 
No. DE-AC02-06CH11357.

\end{acknowledgments}

\bibliography{paper}

\end{document}